\documentclass[conference, twocolumn]{IEEEConf}
\pdfoutput=1
\usepackage[T1]{fontenc}
\usepackage[latin9]{inputenc}
\usepackage{float}
\usepackage{amsmath}
\usepackage{amsthm}
\usepackage{amssymb}
\usepackage{graphicx}

\makeatletter
\theoremstyle{definition}
\newtheorem{assumption}{Assumption}
\theoremstyle{remark}
\newtheorem{rem}{\protect\remarkname}

\usepackage{cite}
\IEEEoverridecommandlockouts

\makeatother

\providecommand{\remarkname}{Remark}

\begin{document}

\title{\bstctlcite{IEEEexample:BSTcontrol}Estimation and Tracking of a
Moving Target by Unmanned Aerial Vehicles \thanks{$^{1}$Department of Mechanical Engineering, National Chiao Tung University,
Hsinchu, Taiwan 30010 Email: michael1874888@gmail.com, amyking90511@gmail.com,
tenghu@g2.nctu.edu.tw}\thanks{This research is supported by the Ministry of Science and Technology,
Taiwan (Grant Number 107-2628-E-009-005-MY3, 106-2622-8-009-017),
and partially supported by Pervasive Artificial Intelligence Research
(PAIR) Labs, Taiwan (Grant Number MOST 108-2634-F-009-006-).}}

\author{Jun-Ming Li$^{1},$ Ching Wen Chen$^{1},$ and Teng-Hu Cheng$^{1}$}
\maketitle
\begin{abstract}
An image-based control strategy along with estimation of target motion
is developed to track dynamic targets without motion constraints.
To the best of our knowledge, this is the first work that utilizes
a bounding box as image features for tracking control and estimation
of dynamic target without motion constraint. The features generated
from a You-Only-Look-Once (YOLO) deep neural network can relax the
assumption of continuous availability of the feature points in most
literature and minimize the gap for applications. The challenges are
that the motion pattern of the target is unknown and modeling its
dynamics is infeasible. To resolve these issues, the dynamics of the
target is modeled by a constant-velocity model and is employed as
a process model in the Unscented Kalman Filter (UKF), but process
noise is uncertain and sensitive to system instability. To ensure
convergence of the estimate error, the noise covariance matrix is
estimated according to history data within a moving window. The estimated
motion from the UKF is implemented as a feedforward term in the developed
controller, so that tracking performance is enhanced. Simulations
are demonstrated to verify the efficacy of the developed estimator
and controller.
\end{abstract}


\section{Introduction}

Knowledge about the position and velocity of surrounding objects is
important to the booming fields such as self-driving cars, target
tracking and monitoring. In case of performing an object tracking
task, position and velocity of the tracking target are typically assumed
to be available to achieve better control performance \cite{Gans2008c}
and \cite{Dani2009} using visual servo controllers. When the target
is not static, its velocity needs be considered in the system dynamics
as to eliminate the tracking error and to calculate the accurate motion
command for the camera. However, obtaining the knowledge online is
challenging since the dynamics of the target might be complicated
and unknown. Moreover, there are instances that the measurement can
be unexpected. For example, the target can exceed the field of view
(FOV) of the camera, or cannot be detected due to the unexpected occlusion.
Several approaches have been proposed for estimating position or velocity
of the target such as by using a fixed camera\cite{Chitrakaran2005},
sensor networks \cite{Stroupe2001,Kamthe2009,Ahn2009}, radar \cite{Schlichenmaier2017},
and some known reference information in the image scene\cite{Huang2017}.
In order to integrate with applications based on vision system such
as target tracking, exploration, visual servo control and navigation
\cite{Gans2008c} and \cite{Thomas2017,Palazzolo2017,Scaramuzza2014},
an algorithm for a monocular camera to estimate position and velocity
of a moving target is developed in this work.

To continuously estimate the position or the velocity of a target,
it needs to remain in the field of view of the camera, and therefore,
motion of the target should be considered. Structure from motion (SfM),
Structure and Motion (SaM) methods are usually used to reconstruct
the relative position and motion between the vision system and objects
in many applications \cite{Gans2008c} and \cite{Dani2009}. With
the knowledge of length between two feature points, \cite{Chitrakaran2007}
proposed methods to estimate position of the stationary features.
In \cite{DawsonRiseRangeID}, 3D Euclidean structure of a static object
is estimated based on the linear and angular velocities of a single
camera mounted on a mobile platform, where the assumption is relaxed
in \cite{Dani2012}. However, SfM can only estimate the position of
the object and usually the object is assumed to be stationary. In
order to address the problem to estimate motion of moving objects,
SaM is applied for estimation by using the knowledge of camera motion.
Nonlinear observers are proposed in \cite{Dani2010b} and \cite{Dani2011}
to estimate the structure of a moving object with time-varying velocities.
The velocity of the object in \cite{Dani2010b} is assumed to be constant,
and \cite{Dani2011} relaxes the constant-velocity assumption to time-varying
velocities for targets moving in a straight line or on a ground plane.
In practice, measurement can be intermittent when the object is occluded,
outside the camera FOV, etc. \cite{Parikh.Cheng.ea2014,Parikh.Cheng.ea2015,Parikh.Cheng.ea2017}
present the development of dwell time conditions to guarantee that
the state estimate error converges to an ultimate bound under intermittent
measurement. In \cite{Parikh.Cheng.ea2014,Parikh.Cheng.ea2015,Parikh.Cheng.ea2017},
the estimation is based on the knowledge about the velocity of the
moving object and the camera. However, in practice the velocity of
the target is usually unknown, and modeling its dynamics is complicated
and challenging.

In fact, the relationship between target motion estimator and vision-based
controller is inseparable. Specifically, output from a  high performance
target motion estimator can be used as a feedforward term for the
controller to keep the target in the field of view longer, which,
in return, results in a longer period for the estimate error to converge.
In this work, a dynamic monocular camera is employed to estimate the
position and velocity of a moving target. Compared to the multi-camera
system\cite{Zhang2016}, using a monocular camera has the advantage
of reducing power consumption and the quantity of image data. A You-Only-Look-Once
(YOLO) deep neural network\cite{Redmon2018} is applied in this work
for target detection, which relaxes the assumption of continuous availability
of the feature point and minimizes the gap for applications, but it
also introduces some challenges. That is, the detected box enclosing
the target can lead to intermittent measurement, and the probability
distribution function of the noise from inaccurate motion model may
not follow the normal distribution. An Unscented Kalman Filter (UKF)
based algorithm is developed in this work to deal with problems of
intermittent measurement and to obtain continuous estimate the target
motion even when it leaves the FOV. To deal with the uncertain noise
during the estimation, method in \cite{Gao2015} is applied to update
the process noise covariance matrix online to guarantee the convergence
and the accuracy of the estimation.

\section{Preliminaries and Problem Statement\label{sec2:Preliminaries-and-problem}}

\subsection{Kinematics Model\label{subsec:Kinematics-Model}}

\begin{figure}[H]
\centering{}\includegraphics[width=1\columnwidth]{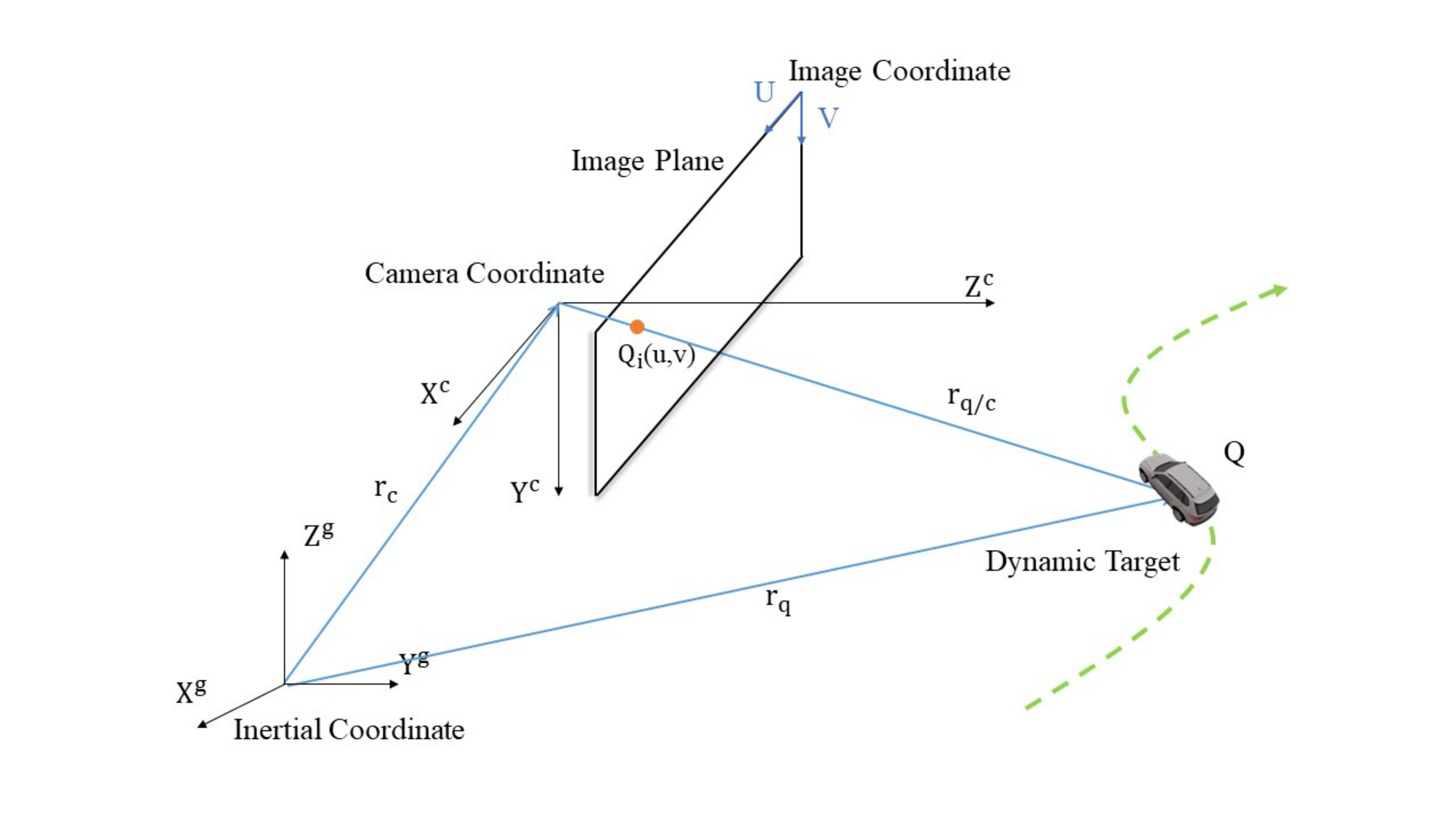}\caption{Kinematics model.\label{fig:model}}
\end{figure}

Based on the model in \cite{Parikh.Cheng.ea2017}, Fig. \ref{fig:model}
depicts the relationship between a moving target, a camera, and an
image plane. The camera is mounted on the multirotor without relative
motion. The subscript $\mathcal{G}$ denotes the inertial frame with
its origin set arbitrarily on the ground, and the subscript $C$ represents
the body-fixed camera frame with its origin fixed at the principle
point of the camera, where $Z^{c}$ and $X^{c}$ are axes with denoted
direction. The vectors $r_{q}=\left[\begin{array}{ccc}
x_{q} & y_{q} & z_{q}\end{array}\right]^{T}$ denotes the position of the feature point of the target, which is
unknown and to be estimated, $r_{c}=\left[\begin{array}{ccc}
x_{c} & y_{c} & z_{c}\end{array}\right]^{T}$ denotes the position of the camera, which can be measured by the
embedded GPS/Motion Capture Systems, and $r_{q/c}=\left[\begin{array}{ccc}
X & Y & Z\end{array}\right]^{T}$ denotes the relative position between the feature point and the camera,
all expressed in the camera frame. Their relation can be written as
\begin{eqnarray}
r_{q/c} & = & r_{q}-r_{c}.\label{eq:rqc}
\end{eqnarray}
Taking the time derivative on the both sides of (\ref{eq:rqc}) yields
the relative velocity as
\begin{eqnarray}
\dot{r}_{q/c} & = & V_{q}-V_{c}-\omega_{c}\times r_{q/c},\label{eq:Vqc}
\end{eqnarray}
where $V_{c}\triangleq\left[\begin{array}{ccc}
v_{cx} & v_{cy} & v_{cz}\end{array}\right]^{T}$ is the linear velocity of the camera, $\omega_{c}\triangleq\left[\begin{array}{ccc}
\omega_{cx} & \omega_{cy} & \omega_{cz}\end{array}\right]^{T}$ is the angular velocity of the camera, both are the control command
to be designed. In (\ref{eq:Vqc}), $V_{q}=\left[\begin{array}{ccc}
v_{qx} & v_{qy} & v_{qz}\end{array}\right]^{T}$ is the linear velocity of the dynamic target, which is unknown and
needs to be estimated. To relax the limitation of existing results,
following assumption is made throughout this work.
\begin{assumption}
\label{assumption:velocity change-1}The trajectory of the target
is unknown but bounded.
\end{assumption}
Since the dynamics of the camera and the target are coupled, the states
of the overall system are defined as 
\begin{equation}
\mathbf{x}=\left[\begin{array}{ccccccccc}
x_{1} & x_{2} & x_{3} & x_{q} & y_{q} & z_{q} & v_{qx} & v_{qy} & v_{qz}\end{array}\right]^{T}.\label{eq:system state}
\end{equation}
 To estimate the position and velocity of the target, the state $\left[\begin{array}{ccc}
x_{1} & x_{2} & x_{3}\end{array}\right]^{T}=\left[\begin{array}{ccc}
\frac{X}{Z} & \frac{Y}{Z} & \frac{1}{Z}\end{array}\right]^{T}$ is defined to facilitate the subsequent analysis. Taking the time
derivative on the both sides of (\ref{eq:system state}) and using
(\ref{eq:Vqc}) obtain a nonlinear function that represents the dynamics
of the overall system as

\begin{eqnarray}
\dot{\mathbf{x}} & = & \left[\begin{array}{c}
v_{qx}x_{3}-v_{qz}x_{1}x_{3}+\zeta_{1}+\eta_{1}\\
v_{qy}x_{3}-v_{qz}x_{2}x_{3}+\zeta_{2}+\eta_{2}\\
-v_{qz}x_{3}^{2}+v_{cz}x_{3}^{2}-(\omega_{cy}x_{1}-\omega_{cx}x_{2})x_{3}\\
V_{q}\\
\begin{array}{c}
0\\
0\\
0
\end{array}
\end{array}\right],\label{eq:dynamic}
\end{eqnarray}
where $\zeta_{1}$, $\zeta_{2}$, $\eta_{1}$, $\eta_{2}\in\mathbb{R}$
are defined as
\begin{align}
\zeta_{1} & =\omega_{cz}x_{2}-\omega_{cy}-\omega_{cy}x_{1}^{2}+\omega_{cx}x_{1}x_{2}\nonumber \\
\zeta_{2} & =-\omega_{cz}x_{1}+\omega_{cx}+\omega_{cx}x_{2}^{2}-\omega_{cy}x_{1}x_{2}\nonumber \\
\eta_{1} & =(v_{cz}x_{1}-v_{cx})x_{3}\nonumber \\
\eta_{2} & =(v_{cz}x_{2}-v_{cy})x_{3}.\label{eq:simplify}
\end{align}

\begin{rem}
Since the trajectory and motion pattern of the target is unknown,
it is modeled by a zero acceleration (i.e., constant velocity) dynamics
as formulated in (\ref{eq:dynamic}), which is reasonable during a
short sampling time with the unneglectable mass of the moving target.
The mismatch between the true and modeled dynamics can be considered
as a process noise in an UKF developed in the subsequent section. 
\end{rem}

\subsection{Image Model\label{subsec:Image-Model}}

\begin{figure}[H]
\centering{}\includegraphics[width=0.48\columnwidth]{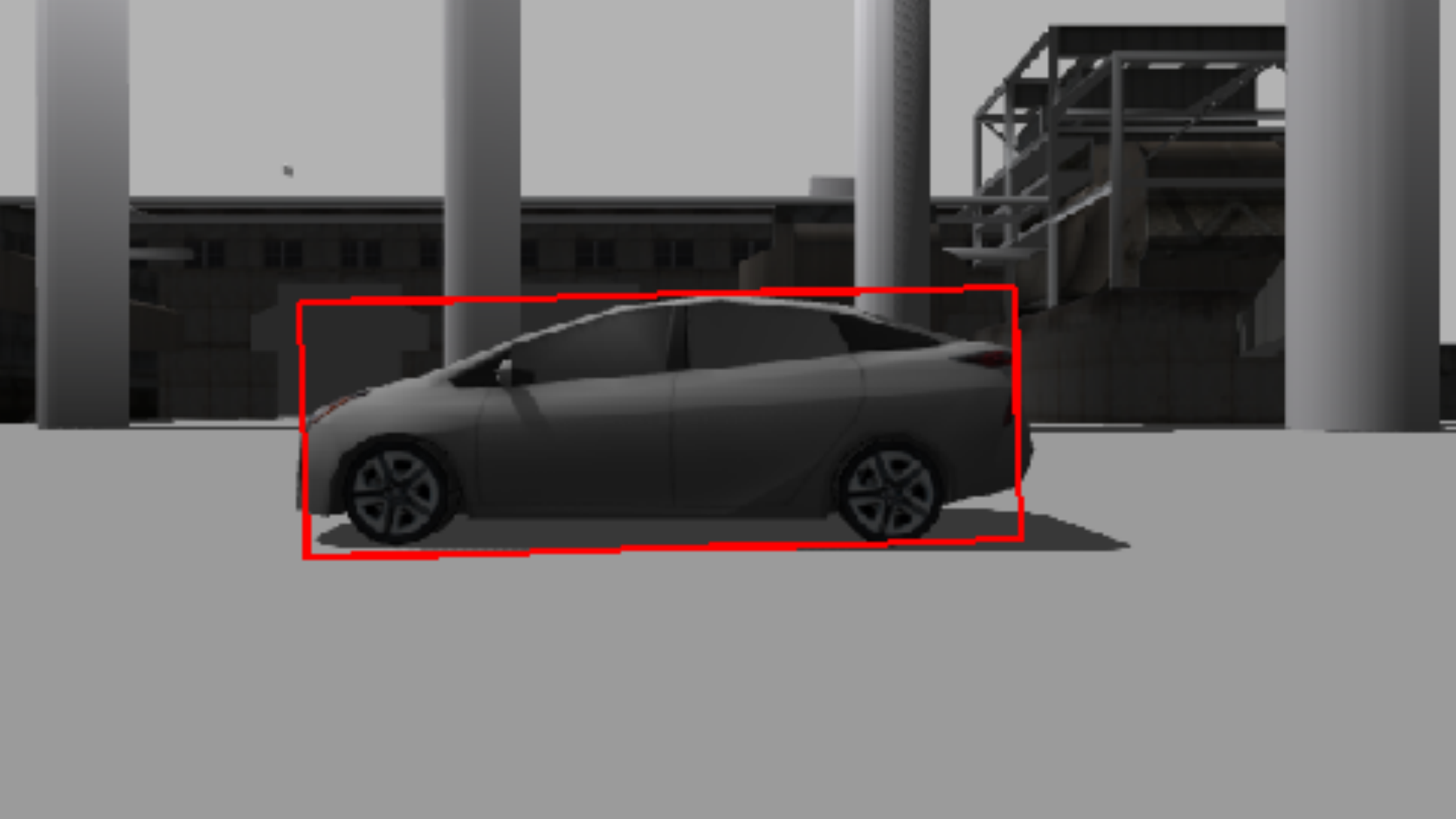}
\includegraphics[width=0.48\columnwidth]{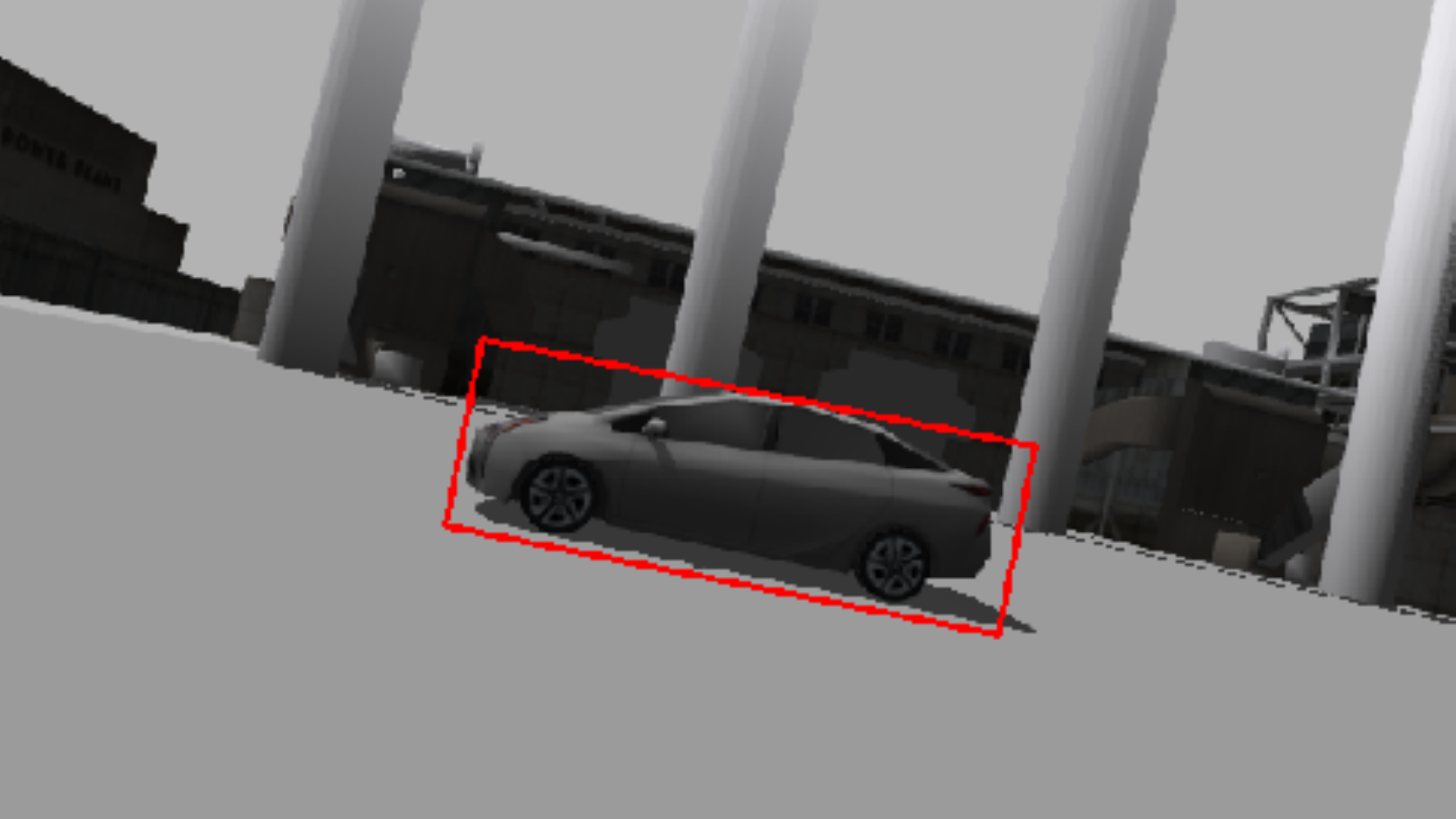}\caption{The images of the dynamic targets are captured from an onboard camera
on the multirotor in the Gazebo simulator. Note that the center of
the bounding box is considered as a feature point for the subsequent
analysis, and the bounding boxes, enclosing the vehicles from different
angles of inclination, are obtained from a YOLO network that is trained
for this work. \label{fig:onboard}}
\end{figure}

By projecting the feature point $Q$ into the image frame using the
pinhole model yields the projection point $q=\left[\begin{array}{cc}
x_{1} & x_{2}\end{array}\right]^{T}\in\mathbb{R}^{2}$ as

\begin{equation}
\begin{array}{c}
x_{1}=\frac{u-c_{u}}{f_{x}}\\
x_{2}=\frac{v-c_{v}}{f_{y}},
\end{array}\label{eq:imModel}
\end{equation}
where $[u,\,v]^{T}$ denotes the position of the feature point in
the image frame, $f_{x}$ and $f_{y}$ are the focal length of pixel
unit, and $[c_{u},\,c_{v}]^{T}$ represents the position of the center
of the image. The area of the bounding box is defined as $a$ , and
based on the pinhole model the relation between $a$ and $x_{3}$
can be expressed as
\begin{eqnarray}
a & = & Af_{x}f_{y}x_{3}^{2}\label{eq:area}
\end{eqnarray}
where $A$ is the area of the target on the side, observed from the
camera\footnote{Given a sedan as the target, the $A$ is about $4.6$m $\times$ 1.5m.}. 
\begin{assumption}
\label{assumption2}The optical axis of the camera remains perpendicular
to $A$ to ensure better detection accuracy from YOLO.
\end{assumption}
\begin{rem}
To estimate $x_{3}$ precisely from (\ref{eq:area}), $A$ needs to
be accurate. Since $A$ is a fixed value, the optical axis of the
camera needs to remain at fixed angle relative to the plane of $A.$
\end{rem}

\subsection{Measurement Model}

To correct the unobserved system states, the measurement is defined
as 
\[
\mathbf{z}=\left[\begin{array}{cccc}
u & v & a & r_{c}\end{array}\right]^{T},
\]
 where $u$, $v$ and $a$ can be obtained directly from the detected
bounding box, and $r_{c}$ is measurable as described in Section \ref{subsec:Kinematics-Model}.
By using (\ref{eq:rqc}), (\ref{eq:imModel}), and (\ref{eq:area}),
the estimate measurement $\hat{\mathbf{z}}$ for the UKF can be obtained
as
\begin{equation}
\hat{\mathbf{z}}=\left[\begin{array}{c}
f_{x}\hat{x}_{1}+c_{u}\\
f_{y}\hat{x}_{2}+c_{v}\\
Af_{x}f_{y}\hat{x}_{3}^{2}\text{sgn}(\hat{x}_{3})\\
\hat{r}_{q}-\hat{r}_{q/c}
\end{array}\right]\label{eq:measurement}
\end{equation}
where $\text{sgn}(\cdotp)$ is a signum function, and $\hat{\left(\cdot\right)}$
is the estimate of the denoted argument obtained from the process
step in the UKF developed in the next section. In (\ref{eq:area}),
the area of bounding box $a$ remains positive despite of the sign
of $x_{3},$ which is positive since the depth is nonnegative. Therefore,
to ensure $\hat{x}_{3}$ converge to a positive value, the term $\text{sgn}(\hat{x}_{3})$
is added to (\ref{eq:measurement}).
\begin{rem}
\label{rem:Despite-the-aforementioned}Despite the aforementioned
advantages, bounding boxes can lose unexpectedly, or the Intersection
over Union (IoU) may sometime decrease, leading to intermittent or
inaccurate measurements. These inherited defects from the data-driven-based
detection motivate the need of Kalman filter for estimation. As the
target velocity changes, state predicted by the constant-velocity
dynamics model can be inaccurate, and the prediction error can considered
as process noise.
\end{rem}

\section{Position and Velocity Estimation\label{sec:Position-and-Velocity}}

\subsection{\label{subsec:UKF}Unscented Kalman Filter}

To estimate state of dynamic systems with noisy measurement or intermittent
measurement, Unscented Kalman Filter\cite{EricA.Wan2000} has been
applied in this work. Based on (\ref{eq:dynamic}) and (\ref{eq:measurement}),
the UKF for nonlinear dynamic system can be expressed as
\begin{align}
\mathbf{x}_{k+1} & =F(\mathbf{x}_{k})+w_{k},\label{eq:process}\\
\mathbf{z}_{k} & =H(\mathbf{x}_{k})+v_{k},\label{eq:measurement-UKF}
\end{align}
where $w_{k}$ and $v_{k}$ represent the process and measurement
noise, respectively, and $F(\cdot)$ and $H(\cdot)$ are the corresponding
nonlinear dynamics and measurement model defined in (\ref{eq:dynamic})
and (\ref{eq:measurement}), respectively.

Based on Remark \ref{rem:Despite-the-aforementioned}, the YOLO detection
might fail incidentally, which makes the measurement correction step
in (\ref{eq:measurement-UKF}) unavailable. When it happens, the state
is only predicted by the dynamics model using (\ref{eq:process}),
which is used as a feedforward term to keep the target in the field
of view, which is reliable in a short period of time before the detection
is recovered. 

\subsection{\label{subsec:Estimation-of-Covariance}Estimation of Noise Covariance
Matrices}

When applying Kalman filter, the process and measurement noise covariance
matrices are usually provided in prior. As mentioned in Remark \ref{rem:Despite-the-aforementioned},
the unmodeled dynamics model can be considered as process noises,
and the covariance matrix is sensitive to the convergence of estimation.
It has been confirmed in our simulations that inaccurate constant
covariance matrices can lead to large estimate error or converge failure.
To dynamically estimate the process noise covariance matrices, a method
developed in \cite{Gao2015} is applied in this work to estimate and
update the covariance matrices online, so that a faster and reliable
convergence performance can be obtained. That is, the process noise
$w_{k}$ is assumed to be uncorrelated, time-varying, and nonzero
means Gaussian white noises that satisfies
\begin{align}
Q_{k}\delta_{kj} & =cov(w_{k},w_{j})\label{eq:QRdefinition}
\end{align}
where $\delta_{kj}$ is the Kronecker $\delta$ function. By selecting
a window of size $N,$ the estimate of the process noise covariance
matrix $\hat{Q}_{k-1}\in\mathbb{R}^{m\times m}$ can be expressed
as
\begin{align}
\hat{Q}_{k-1} & ={\displaystyle \sum_{j=1}^{N}v_{j}\left[P_{k-j}+K_{k-j}\varepsilon_{k-j}\varepsilon_{k-j}^{T}K_{k-j}^{T}\right.}\label{eq:Q_estimate}\\
 & -\sum_{i=0}^{2n}\omega_{i}^{c}\left(\xi_{i,k-j/k-1-j}-\hat{X}_{k-j/k-1-j}\right)\times\nonumber \\
 & \left.\left(\xi_{i,k-j/k-1-j}-\hat{X}_{k-j/k-1-j}\right)^{T}\right],\nonumber 
\end{align}
Since $\hat{Q}_{k-1}$ might not be a diagonal matrix and positive
definite, it is further converted to a diagonal, positive definite
matrix as
\begin{equation}
\hat{Q}_{k-1}^{*}=\text{diag}\left\{ |\hat{Q}_{k-1}(1)|,|\hat{Q}_{k-1}(2)|,\cdots,|\hat{Q}_{k-1}(m)|\right\} ,\label{eq:Qestimate}
\end{equation}
where $\hat{Q}_{k-1}(i)$ is the $i$-th diagonal element of the matrix
$\hat{Q}_{k-1}$. On the other hand, the measurement noise can be
measured in advance.

\section{Tracking Control}

In this section, a motion controller for the multirotor is designed
using vision feedback. Compared to the existing Image-based Visual
Servo (IBVS) control methods\cite{Chaumette2006I}, the controller
developed in this work not only uses feedback but also includes a
feedforward term to compensate the target motion and to ensure better
tracking performance, where the feedforward term is obtained from
the UKF developed in Section \ref{subsec:UKF}. Most existing approaches
either focus on the estimate of target position/velocity or camera
position/velocity, but yet the controllers designed for the cameras
are rarely discussed, and vice versa. Additionally, the relation between
estimating the target motion and controlling the camera are highly
coupled. That is, a high performance motion controller can minimizes
the estimate error (i.e., the camera is controlled to keep the target
in the field-of-view longer), which, in return, yields a precise feedforward
term to facilitate the tracking performance, and vice versa. Finally,
since YOLO deep neural network is employed to enclose the target in
the image, the envelop area is defined as a new reference signal for
the controller to track. 

\subsection{Target Recognition }

YOLO\cite{Redmon2018} is a real-time object detection system with
reasonable accuracy after training. Our YOLO network is trained by
using a large number of dataset and the performance is verified before
implementation in this work. 

\subsection{Controller\label{subsec:Controller}}

The IBVS controller based on \cite{Shahriari2013} is employed in
this work for achieving tracking control of dynamic targets. To this
end, a vector $s\left(t\right)=\left[x_{1},\,x_{2},\,x_{3}\right]^{T}\colon\left[0,\,\infty\right)\rightarrow\mathbb{R}^{3}$
denoted a feature vector is defined as the control state which is
defined in (\ref{eq:system state}). The visual error $e(t)\colon\left[0,\,\infty\right)\rightarrow\mathbb{R}^{3}$
to be controlled is defined as

\begin{equation}
e=s-s^{*}\label{eq:controlError}
\end{equation}
where $s^{*}\in\mathbb{R}^{3}$ is a desired constant vector of the
feature vector predefined by the user (i.e., typically $\left[x_{1}^{*},\,x_{2}^{*}\right]^{T}$
is selected as the center of the image and $x_{3}^{*}$ is a function
of the expected distance to the target). Taking the time derivative
of (\ref{eq:controlError}) and using (\ref{eq:dynamic}) yield the
open-loop error system as
\[
\dot{e}=\dot{s}=L_{e}\left[\begin{array}{c}
V_{c}-V_{q}\\
\omega_{c}
\end{array}\right],
\]
where $V_{c}$ and $\omega_{c}$ are considered as the control inputs,
$V_{q}$ is the feedforward term estimated by the UKF, and $L_{e}\in\mathbb{R}^{3\times6}$
is the interaction matrix defined as
\begin{align}
L_{e} & =\nonumber \\
 & \left[\begin{array}{cccccc}
-x_{3} & 0 & x_{1}x_{3} & x_{1}x_{2} & -\left(x_{1}^{2}+1\right) & x_{2}\\
0 & -x_{3} & x_{2}x_{3} & (x_{2}^{2}+1) & -x_{1}x_{2} & -x_{1}\\
0 & 0 & x_{3}^{2} & x_{2}x_{3} & -x_{1}x_{3} & 0
\end{array}\right].\label{eq:Le}
\end{align}

\begin{figure}
\centering{}\includegraphics[width=0.99\columnwidth]{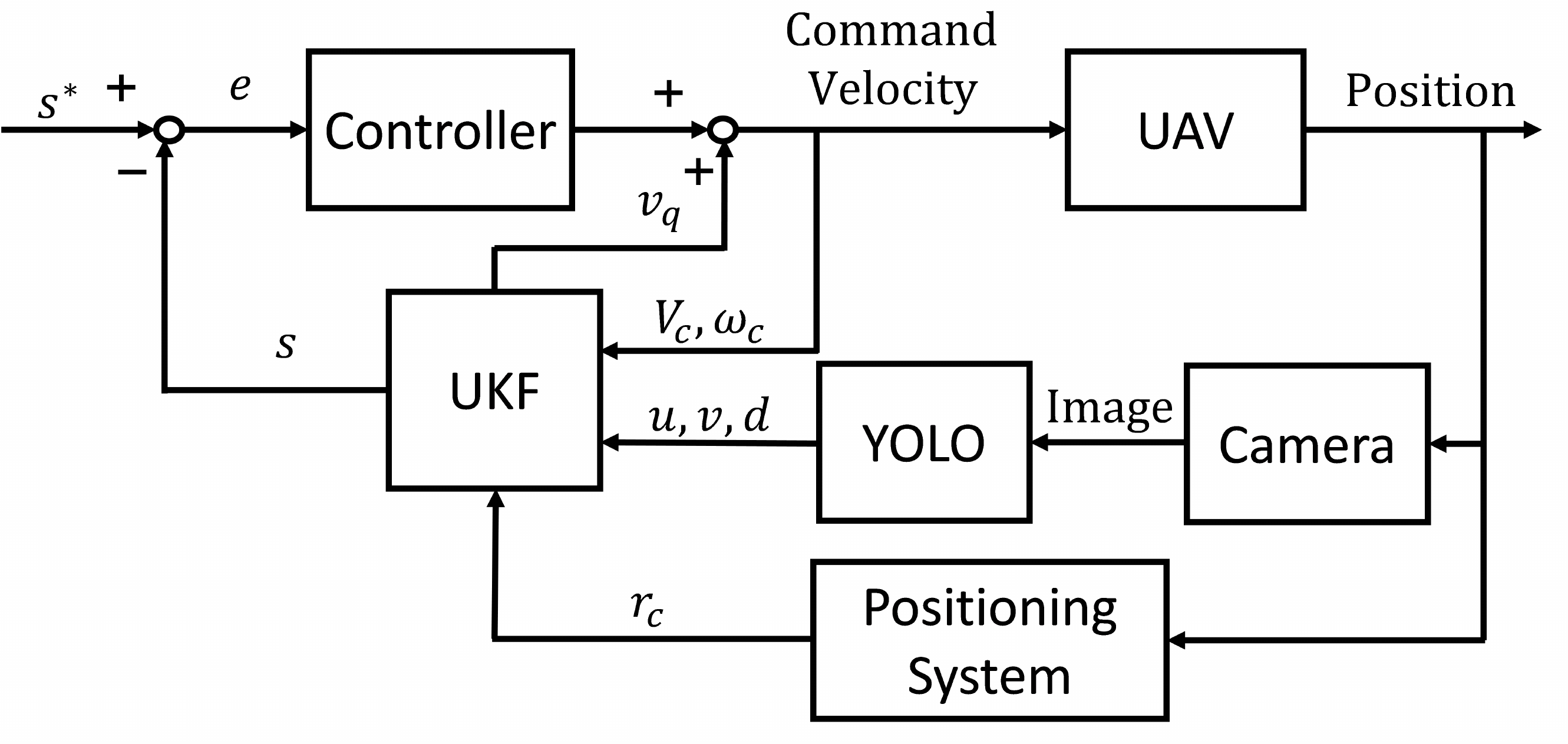}\caption{Block diagram of the controller. \label{fig:Block-Diagram}}
\end{figure}

Note that as the error signal $e$ converges to zero, the position
of the camera relative to the target is not unique, due to the fact
that the camera control input $V_{c}$ and $\omega_{c}$ have a higher
dimension compared to $e.$ To keep the camera staying on the left-hand-side
of the target as to maintain high detection accuracy from YOLO\footnote{Pose estimate of the target at this angle can be achieved by a well-trained
YOLO, and the extension to multiple angles of view will be trained
in the future.}, $v_{cx}$ is controlled to track the moving target as to stay on
the specified angle facing toward the target as
\begin{equation}
v_{cx}=-\frac{w_{im}d_{exp}(\psi-\psi_{exp})}{FOV_{u}\times f_{x}},\label{eq:vcx}
\end{equation}
where the design is inspired from \cite{Pestana2014}. In (\ref{eq:vcx}),
$w_{im}$ is the width of the image in pixel, $d_{exp}$ denotes the
expected distance to the target, $FOV_{u}$ is the horizontal field
of view of the camera, and $\psi$ and $\psi_{exp}$ are the current
and the expected angle of view with respect to the target, respectively.
Since $v_{cx}$ is specified in (\ref{eq:vcx}), the corresponding
column in the interaction matrix $L_{e}$ defined in (\ref{eq:Le})
can be removed, which gives the resultant matrix $\hat{L}_{e}\in\mathbb{R}^{3\times5}$
as
\begin{align}
\hat{L}_{e} & =\nonumber \\
 & \left[\begin{array}{ccccc}
0 & x_{1}x_{3} & x_{1}x_{2} & -\left(x_{1}^{2}+1\right) & x_{2}\\
-x_{3} & x_{2}x_{3} & (x_{2}^{2}+1) & -x_{1}x_{2} & -x_{1}\\
0 & x_{3}^{2} & x_{2}x_{3} & -x_{1}x_{3} & 0
\end{array}\right].\label{eq:Le_m}
\end{align}
Using the Moore-Penrose pseudo-inverse of $\hat{L}_{e}$ as well as
adding a feedforward term, the tracking controller for the camera
can be designed as
\begin{align*}
v_{cx} & =-\frac{w_{im}d_{exp}(\psi-\psi_{exp})}{FOV_{u}\times f_{x}}+v_{qx}\\
\left[\begin{array}{c}
v_{cy}\\
v_{cz}\\
\omega_{cx}\\
\omega_{cy}\\
\omega_{cz}
\end{array}\right] & =-\lambda\hat{L}_{e}^{+}e+\left[\begin{array}{c}
v_{qy}\\
v_{qz}\\
0\\
0\\
0
\end{array}\right].
\end{align*}
The block diagram of the controller is shown in Fig. \ref{fig:Block-Diagram}. 

\section{Simulations}

\subsection{Environment Setup}

In the simulation\footnote{https://goo.gl/93EDnd}, a car is considered
as a moving target and is tracked by a quadrotor, where the developed
controller as well as the UKF are implemented. A camera is implemented
on the quadrotor to provide visual feedback. Specifically, bounding
boxes are generated in the image to enclose detected cars as shown
in Fig. \ref{fig:onboard}, which is achieved by a pretrained YOLO
deep neural network. The simulation is conducted in the ROS framework
(16.04, kinetic) with Gazebo simulator. In the simulation environment,
the value of $A=4.6\text{m}\times1.5\text{m},$ and the resolution
of the image is $640$x$480$ with 50 fps. The intrinsic parameters
matrix of the camera is 
\begin{eqnarray*}
K & = & \left[\begin{array}{ccc}
381.36 & 0 & 320.5\\
0 & 381.36 & 240.5\\
0 & 0 & 1
\end{array}\right].
\end{eqnarray*}
which is obtained by calibration.

A time moving window of width $N$ is set to be 150 with sampling
rate of $50\:\text{sample}/\text{sec}$. The initial process and measurement
noise covariance matrices are selected as $diag\left\{ \left[20,\,20,\,500,\,0.0001,\,0.0001,\,0.0001\right]\right\} $
and $diag\left\{ \left[0.08,\,0.08,\,0.02,\,5,\,5,\,5,\,1,\,1,\,1\right]\times10^{-2}\right\} ,$
respectively, and the process noise covariance matrix is estimated
online using (\ref{eq:Qestimate}). 

The initial location of the car and the drone are $\left[\begin{array}{ccc}
0 & 0 & 0\end{array}\right]^{T}$ and $\left[\begin{array}{ccc}
0 & 5.5 & 1.0\end{array}\right]^{T}$ along with the initial orientations $\left[\begin{array}{ccc}
0 & 0 & 0\end{array}\right]^{T}$ and $\left[\begin{array}{ccc}
0 & 0 & -\frac{pi}{2}\end{array}\right]^{T}$ in radians, respectively, all expressed in the global frame. The
car is free to move on the $XY$ plane, and its velocity is specified
based on the real-time user command. 

\subsection{Simulation Results}

\begin{figure}[H]
\begin{centering}
\includegraphics[width=1\columnwidth]{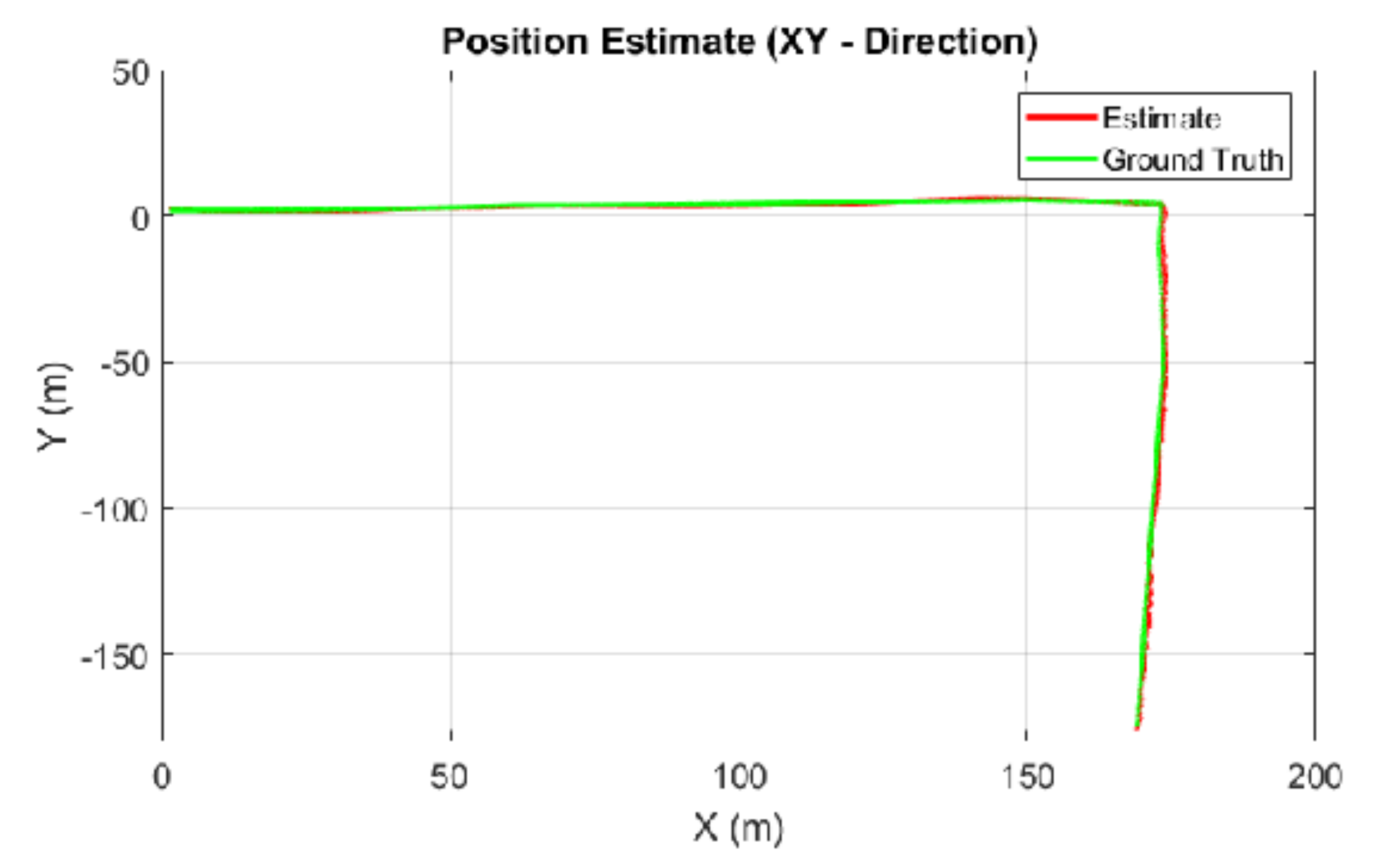}
\par\end{centering}
\begin{centering}
\caption{Estimate and ground truth of the target trajectory in the $XY$ plane
in the global frame.\label{fig:position x}}
\par\end{centering}
\end{figure}

Fig. \ref{fig:position x} depicts the position estimate errors of
the moving vehicle with simulation period of 223 seconds. The position
estimate errors are reduced from 14\% to 7\% as the target moves from
time-varying velocity to constant velocity, despite some noises. Note
that in practice the drone may not react fast enough to the rapid
change of velocity, in which the optical axis cannot remain facing
right to the target, leading to a slight deviation in the depth estimate
(i.e., $x_{3}$). Note that the position estimate error in the $Z$-axis
increases as the velocity in the $Y$-axis increases. This can be
attributed to the fact that the increasing velocity of the vehicle
in the $Y$-axis causes the quadrotor to tilt forward for tracking,
which breaks the assumption \ref{assumption2} that the optical axis
is facing toward the side of the vehicle and leads to a large estimate
error. 
\begin{rem}
In Gazebo environment, the velocity of the car is set below 4 m/s
due to a large drag. As the speed of the car increases, the quadrotor
accelerates with a tilt angle, which increases the chance of object
detection failure. However, the problem can be resolved by expanding
the training dataset with images from different angles of view, which
will be part of the future work. 
\end{rem}

\begin{figure}[H]
\begin{centering}
\includegraphics[width=0.49\columnwidth]{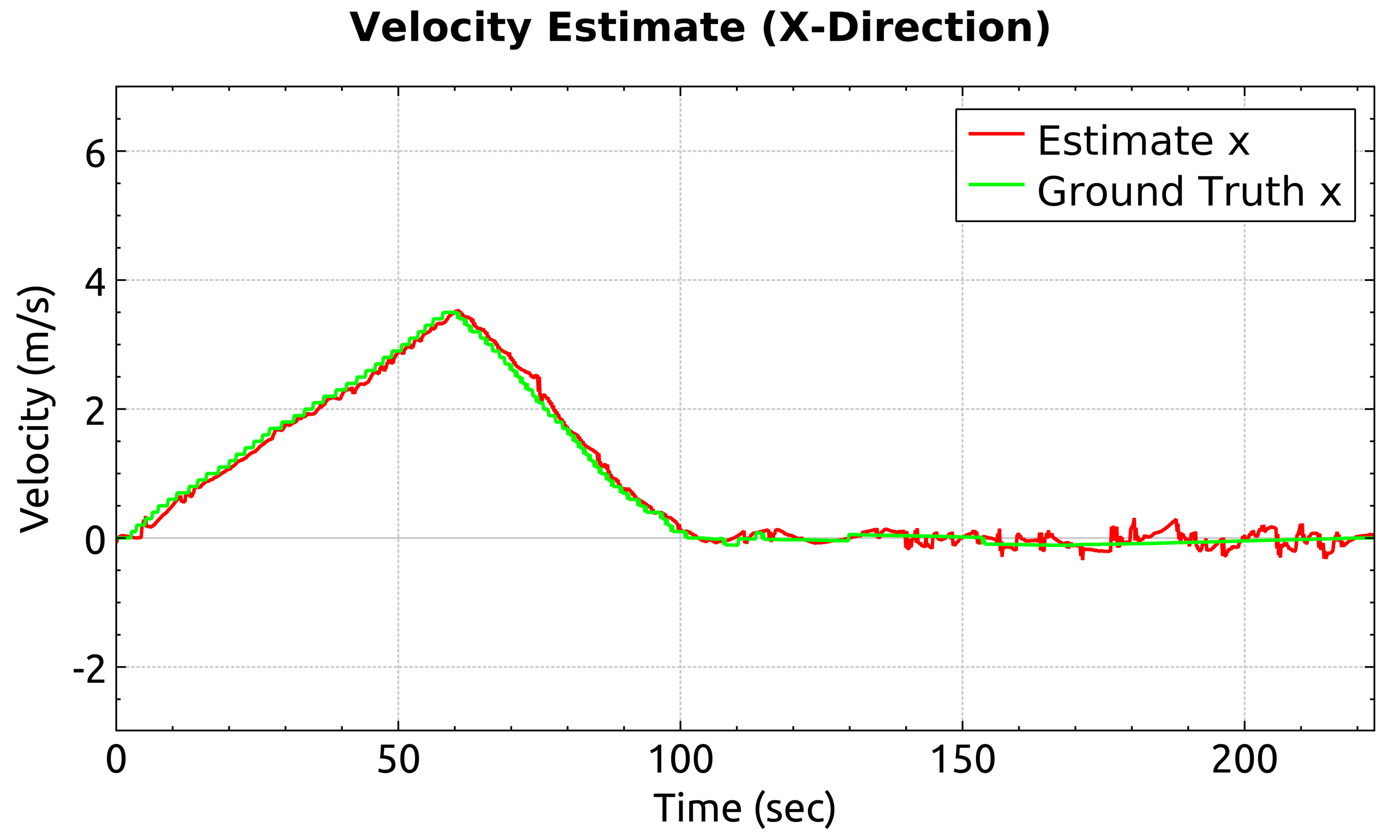}\includegraphics[width=0.49\columnwidth]{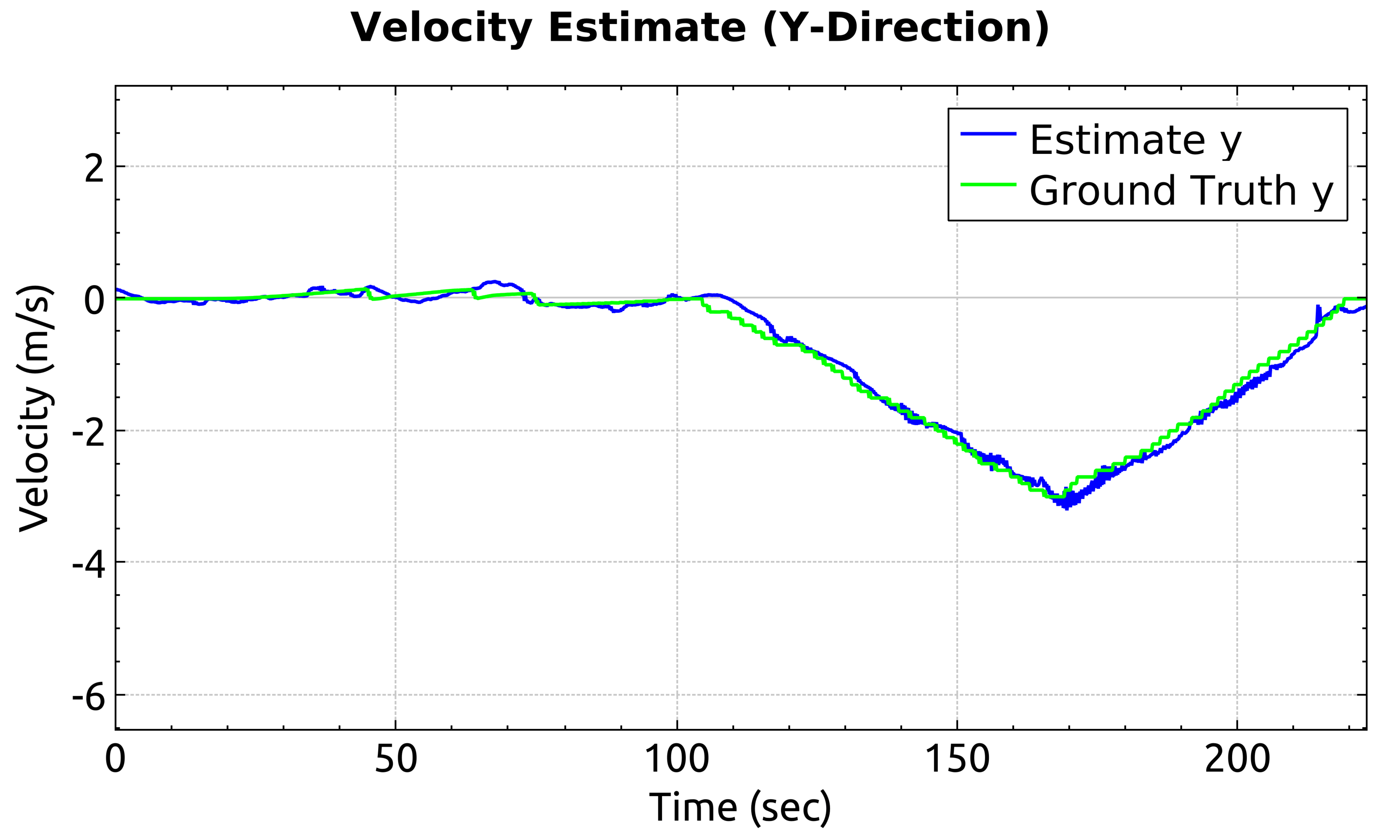}
\par\end{centering}
\centering{}\caption{Estimate and ground truth of the target velocity in the $X$ and $Y$
direction of the global frame.\label{fig:velocity x}}
\end{figure}

Fig. \ref{fig:velocity x} depicts the velocity estimate errors of
the moving vehicle. The estimate performance is slightly compromised
when the vehicle is accelerated (i.e., due to the constant-velocity
model utilized in the UKF), but better estimate performance can be
expected by increasing the sensing rate for the UKF measurement. The
increase of the velocity estimate error between 103-223 seconds can
be attributed to the acceleration of the target in the $Y$-direction.

\begin{figure}[H]

\begin{centering}
\includegraphics[width=0.49\columnwidth]{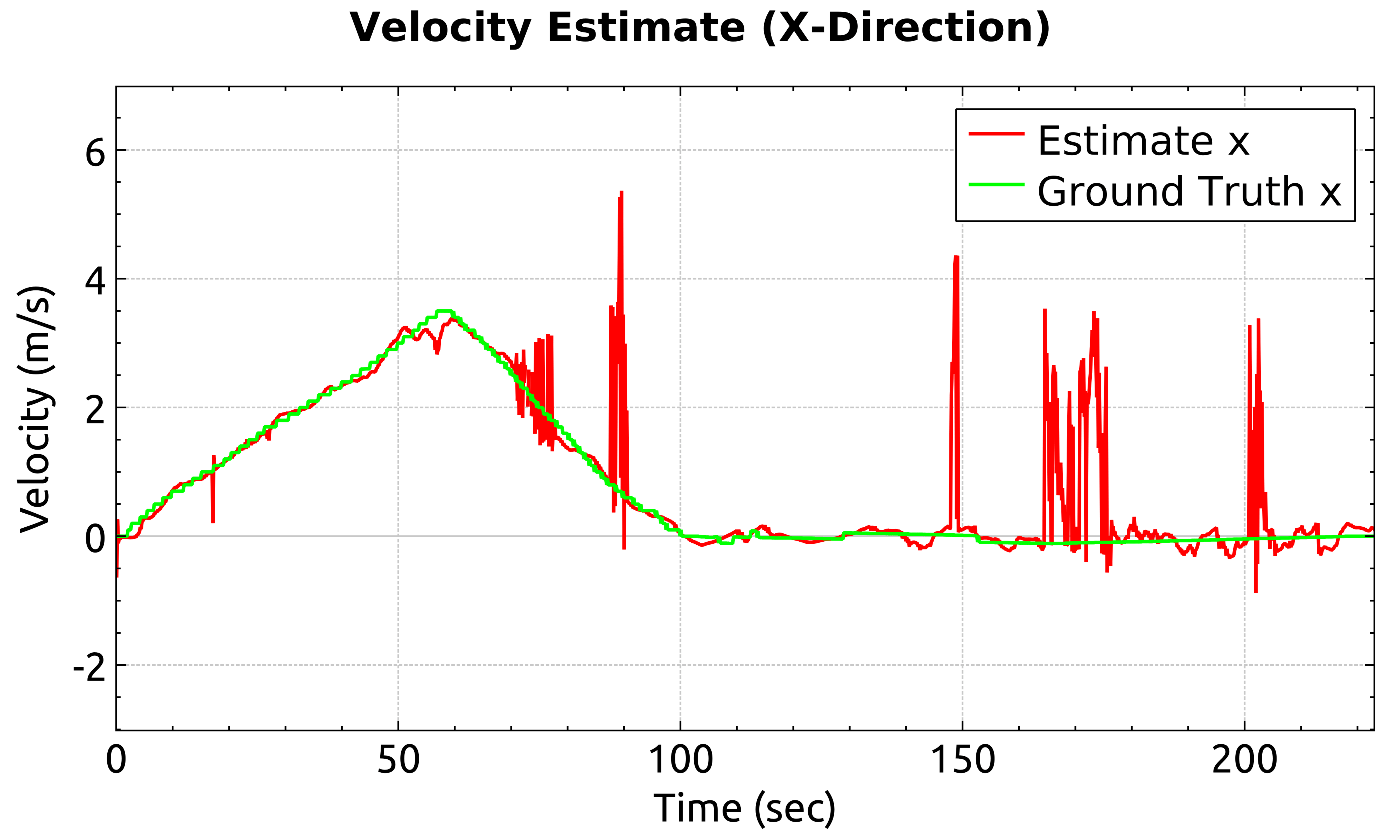}\includegraphics[width=0.49\columnwidth]{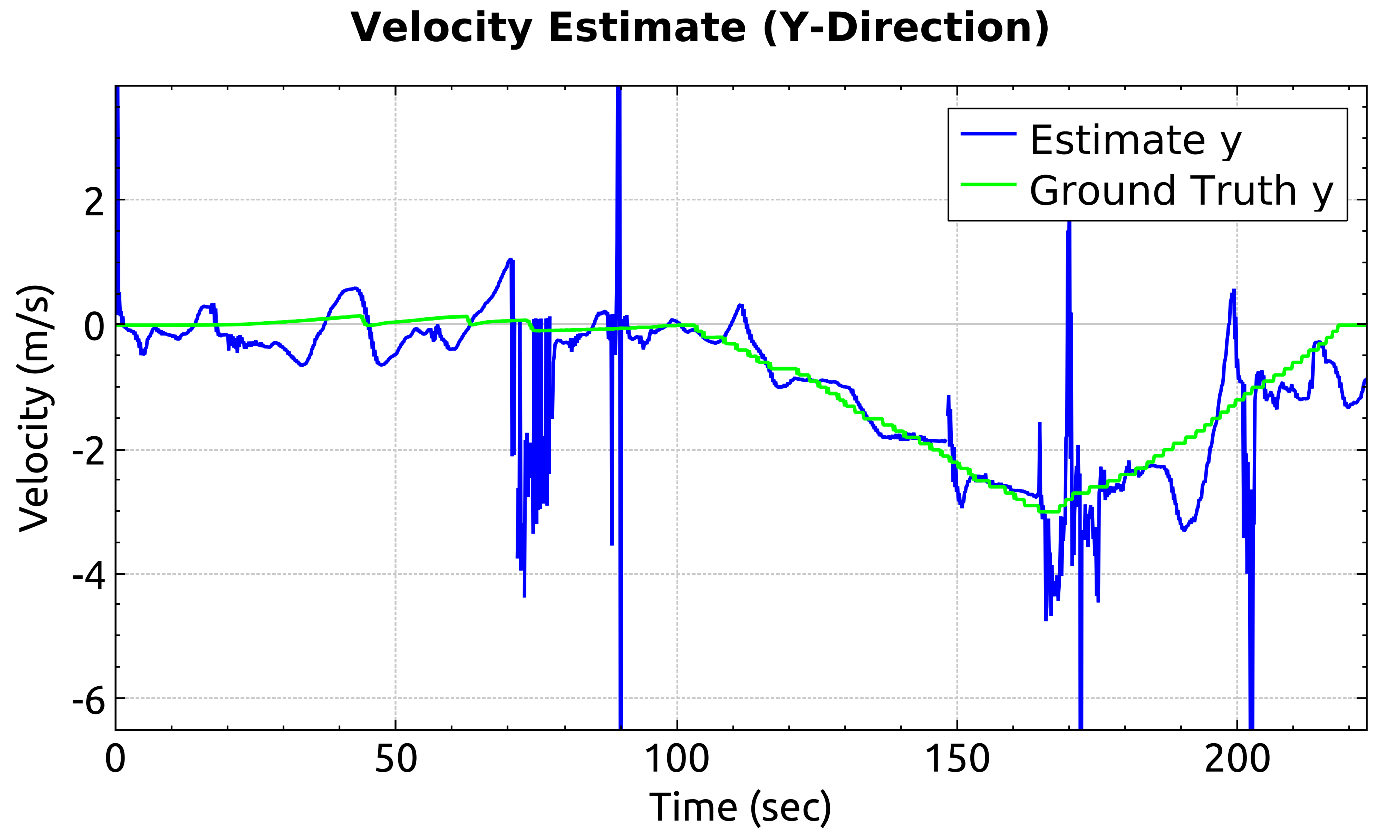}
\par\end{centering}
\caption{Estimate and ground truth of the target velocity without estimation
of noise covariance matrix.\label{fig:With-and-without}}

\end{figure}

Fig. \ref{fig:With-and-without} depicts the estimate error without
process noise estimation, where the constant covariance matrix $Q$
is same as the initial value. Compared to Fig. \ref{fig:velocity x},
using the estimated noise covariance matrix developed in (\ref{eq:Qestimate})
yields a better velocity estimate performance. 

\section{Conclusion}

In this work, a motion controller for a camera on an UAV is developed
to track a dynamic target with unknown motion and without motion constraint.
The unknown target motion is estimated in the developed UKF with process
noise covariance matrix estimated based on the past data within a
moving window, and the intermittent measurement caused by YOLO detection
is addressed. The estimated target velocity is then included as a
feedforward term in the developed controller to improve tracking performance.
Compared to the case without noise estimation, the developed approach
is proven to obtain better tracking performance. Although Assumption
\ref{assumption2} is rigorous in practice, this work is the first
one to prove the feasibility of the overall control architecture,
and future work will be relaxing the assumption by training a YOLO
network to detect the target from any angles. Additionally, eliminating
the knowledge of the ground truth will be another future works.

\bibliographystyle{IEEEtran}
\bibliography{master,ncr,encr}

\end{document}